# Transition-Metal Nitride Halide Dielectrics for Transition-Metal Dichalcogenide Transistors

Mehrdad Rostami Osanloo[1], Ali Saadat[2], Maarten L. Van de Put[2], Akash Laturia[2], and William G. Vandenberghe[2]

[1] Department of Physics, The University of Texas at Dallas, Richardson, TX 75080, USA
[2] Department of Materials Science and Engineering, The University of Texas at Dallas, Richardson, TX 75080, USA

E-mail: william.vandenberghe@utdallas.edu

## Abstract

Using first principles calculations, we investigate six transition-metal nitride halides (TMNHs): HfNBr, HfNCl, TiNBr, TiNCl, ZrNBr, and ZrNCl as potential van der Waals (vdW) dielectrics for transition metal dichalcogenide (TMD) channel transistors. We calculate the exfoliation energies and bulk phonon energies and find that the six TMNHs are exfoliable and thermodynamically stable. We calculate both the optical and static dielectric constants in the in-plane and out-of-plane directions for both monolayer and bulk TMNHs. In monolayers, the out-of-plane static dielectric constant ranges from 5.04 (ZrNCl) to 6.03 (ZrNBr) whereas in-plane dielectric constants range from 13.18 (HfNBr) to 74.52 (TiNCl). We show that bandgap of TMNHs ranges from 1.53 eV (TiNBr) to 3.36 eV (HfNCl) whereas the affinity ranges from 4.01 eV (HfNBr) to 5.60 eV (TiNCl). Finally, we estimate the dielectric leakage current density of transistors with six TMNH monolayer dielectrics with five monolayer channel TMDs ($MoS_2$, $MoSe_2$, $MoTe_2$, $WS_2$, and $WSe_2$). For $p$-MOS TMD channel transistors 19 out of 30 combinations have a smaller leakage current compared to monolayer hexagonal boron nitride (hBN), a well-known vdW dielectric. The smallest monolayer leakage current of $2.14 \times 10^{-9}$ A/cm$^2$ is predicted for a $p$-MOS $WS_2$ transistor with HfNCl as a gate dielectric. HfNBr, HfNCl, ZrNBr, and ZrNCl are also predicted to yield small leakage currents in certain $p$-MOS TMD transistors.

## 1. Introduction

Two-Dimensional (2D) van der Waals (vdW) materials have attracted a huge amount of attention owing to their unique properties and numerous applications in a wide variety of research areas, such as electronics, optoelectronics, and spintronic [1-5]. With the advent of graphene, the first 2D thermodynamically stable material, the electronic industry entered a new area of technological revolution. The search for other promising 2D-vdW materials has led to the discovery of the most reputable class of 2D layered materials, transitional metal dichalcogenides (TMDs), with a chemical formula of $MX_2$ (where M is transition metal and X is a chalcogen) and opened a research avenue for their unique and unprecedented electronic and optical properties [6-8]. It is now well established that monolayers (1L) of TMDs exhibit exceptional physical behaviour compared to their multilayer or bulk form, especially under tensile strain, defect engineering, and/or electric field exposure [9-12]. In this regard, TMDs have been featured as *n*-type and *p*-type channel materials, in combination with 2D dielectric materials, to produce high-performance thin-film transistors (TFTs) [13-15]. Despite several studies performed in this field, there are still numerous issues for transistors with TMD channels and 2D dielectrics which have not yet been properly addressed and required more investigations.

Transition-metal nitride halides (TMNHs) with chemical formula MNX (where M is a metal, N is nitrogen, and X is a halogen) are a less explored type of 2D-vdW material with layered crystal structures bound by a weak vdW force between the layers [16, 17]. Similar to other vdW materials, TMNHs have a huge potential to be utilized in various applications. Among the TMNHs, Group-IVB nitride halides are of great interest since they are environmentally friendly materials [18]. Several studies have been done on layered TMNHs to investigate properties such as superconductivity [19, 20], photocatalyst [18], energy storage capability [21], and electronic properties [22, 23]. Moreover, we can envision materials like HfNCl to be compatible with existing semiconductor technology since $HfCl_4$ is a known Atomic Layer Deposition (ALD) precursor [24, 25] and nitrogen is omnipresent in semiconductor processing. While TMNH electronic transport properties are starting to be explored, another unexplored option is to use TMNHs as vdW dielectrics.

In general, to find a good vdW layered dielectric candidate, we need to consider four criteria: i) Stability: layered dielectric materials must be stable and exfoliable. ii) Leakage current: dielectrics must be a good insulator possessing a reasonable dielectric-channel band offset exceeding 1 eV to diminish leakage current by tunnelling or thermionic emission. iii) Equivalent Oxide thickness (EOT): a small EOT is desirable for the purpose of the device miniaturization [2, 26-30]. iv) Defects: defects introduce new electronic states in the bandgap, such that charge trapped in defects alter the gate threshold voltage. The first three criteria are directly determined by materials properties while the fourth criterion is also heavily determined by experimental process conditions. Despite numerous studies on dielectric properties of vdW materials [2, 31], the investigation for dielectric properties of TMNHs is missing.

In this work, we perform first-principles calculations, using density functional theory (DFT), to study the optical and static in-plane and out-of-plane dielectric properties of monolayer and bulk form of six TMNHs: HfNBr, HfNCl, TiNBr, TiNCl, ZrNBr, and ZrNCl. We calculate the TMNH exfoliation energies and phonon energies to investigate the thermal structural stability of TMNHs. We also calculate bandgaps, electron affinities, and EOTs for the TMNHs under study. Furthermore, we consider five very well-known TMDs ($MoS_2$, $MoSe_2$, $MoTe_2$, $WS_2$, and $WSe_2$) and calculate their band energies by DFT. Finally, we compute the leakage current of *p*-MOS transistors with TMNHs as gate dielectrics and TMDs as channel materials. We find that the combination of TMNH materials under study with introduced TMDs has significant potential for applications in the next generation of *p*-MOS devices.

## 2. Computational methods

To perform our calculations, we obtain our primary input file including atomic positions of materials from Materials Project [32] and use DFT as implemented in the Vienna *ab initio* simulation package (VASP) [33, 34]. For the electron exchange and correlation functional we employ the generalized gradient approximation (GGA) as proposed by Pedrow-Burke-Ernzerof (PBE) [35]. To ensure accuracy we set the plane wave extension energy cut-off of 520 eV for TMNHs, and structural relaxations are performed until the force on each atom is less than $10^{-3}$ eV Å$^{-1}$. To achieve accurate dielectric constants, we set a strict energy convergence criterion of $10^{-8}$ eV. For monolayer TMDs in this paper, we use the plane wave energy cut-off range between 450 eV to 520 eV and set an energy break condition for the electronic self-consistency loop and ionic relaxation loop to $10^{-5}$ eV and $10^{-4}$ eV, respectively. To sample the Brillouin Zone (BZ), 15×15×3 (9×9×9) and 15×15×1 (9×9×1) Gamma k-point grids are adopted for the bulk and the monolayer structures of HfNBr, HfNCl, ZrNBr, ZrNCl (TiNBr, TiNCl), respectively. To account for the interlayer van der Waals interactions, we use the DFT-D3 method of Grimme's [36]. We also apply at least 15 Å vacuum barrier to the monolayer cells to avoid any possible interactions between layers. Furthermore, the exfoliation energy is computed as the ratio between the difference between the ground state energies of bulk and monolayers to the bulks' surface area [37, 38]. To avoid underestimating the band gap, we use the Heyd–Scuseria–Ernzerhof (HSE06) functional in band gap and electron affinity calculations. For monolayer TMD band gaps, we perform HSE06 calculations including spin-orbit coupling (SOC) interactions with 5×5×1 k-point mesh sampling in the first BZ. After determining the ground state on the 5x5x1 mesh, we add two 25-point path along the Γ → M and Γ → K directions to ensure we capture the band minimum/maximum.

To calculate the bulk dielectric constants, we use Density Functional Perturbation Theory (DFPT), implemented in VASP, computing the permittivity tensor of the bulk unit cell. From the permittivity tensor, we derive the in-plane dielectric constant by calculating the geometric mean over the $x$ and $y$ components, $\varepsilon_\parallel = \sqrt{\varepsilon_x \times \varepsilon_y}$, and consider the third component for the out-of-plane dielectric constant, $\varepsilon_\perp = \varepsilon_z$. We calculate the optical dielectric constant ($\varepsilon_\infty$) in high frequency where only the electrons respond to the external field and the ions remain frozen in their positions. To account for the ionic response, we compute the static dielectric constant ($\varepsilon_0$) which includes both the electronic and ionic responses. To analyze the monolayer dielectric constant, we need to remove the vacuum contribution from the supercell dielectric values and rescale the supercell dielectric constants. Following Ref. [2, 31], we rescale the supercell dielectric constants using the following equations:

$$\varepsilon_{2D,\perp} = \left[1 + \frac{c}{t}\left(\frac{1}{\varepsilon_{SC,\perp}} - 1\right)\right]^{-1} \quad (1)$$

$$\varepsilon_{2D,\parallel} = 1 + \frac{c}{t}\left(\varepsilon_{SC,\parallel} - 1\right) \quad (2)$$

where $c$ is the supercell height, and $t$ is the monolayer thickness (shown in **Fig. 1**).

To calculate the EOT, we use

$$EOT = \left(\frac{\varepsilon_{SiO_2}}{\varepsilon_{TMNH}}\right) t_{TMNH}, \quad (3)$$

where $\varepsilon_{SiO_2}$, $\varepsilon_{TMNH}$, and $t_{TMNH}$ are silicon dioxide dielectric constant (3.9), monolayer out-of-plane static dielectric constant of the dielectric, and dielectric monolayer thickness, respectively.

Dielectric leakage current through the metal-semiconductor is calculated by considering three components: thermionic emission over the barriers ($\varphi = [\chi_{TMNH} + E_{g,TMNH}] - [\chi_{TMD} + E_{g,TMD}]$, where $\chi$ is the electron affinity and $E_g$ is the bandgap), tunneling current through the valence band ($\varphi = [\chi_{TMNH} + E_{g,TMNH}] - [\chi_{TMD} + E_{TMD}]$), and tunneling current through the conduction band

($\varphi = [\chi_{\text{TMD}} + E_{\text{TMD}}] - \chi_{\text{TMNH}}$). Fowler-Nordheim equations, as follows, are used to obtain the thermionic and tunneling current densities [39, 40]:

$$J_{\text{tun}} = \frac{q^3 \mathcal{E}^2}{8\pi h \varphi} \exp\left(-\frac{4\sqrt{2m^*}\, \varphi^{3/2}}{3q\hbar\mathcal{E}}\right) \quad (4)$$

$$J_{\text{therm}} = A^{**} T^2 \, \exp\left(\frac{-q\left(\varphi - \sqrt{\frac{q\mathcal{E}}{4\pi\varepsilon_i}}\right)}{kT}\right) \quad (5)$$

where $\mathcal{E}$, $\varphi$, $\varepsilon_i$, $A^{**}$, $T$, $q$, $m^*$, and $k$ are the electric field in the insulator, barrier height, insulator permittivity, effective Richardson constant, temperature, electron charge, electron/hole effective mass, and Boltzmann constant, respectively.

To calculate the leakage current, we use the $V_{DD}$=0.7 V supply voltage and $V_t$=0.345 V threshold voltage as introduced in 2020 International Roadmap for Devices and Systems (IRDS) [41]. The electric field inside the insulator, $\mathcal{E}$, is calculated using $\mathcal{E} = \frac{V_{DD} - V_t}{t}$, where t is the thickness of monolayer TMNHs. Since the band structures of the TMNHs are flat along the out-of-plane direction, we use the free hole mass for the tunneling calculations ($m^* = m_h$). The hBN hole effective mass, 0.47 $m_h$, is taken from Ref. [42]. For the thermionic emission calculation, we use the out-of-plane static dielectric constants. The monolayer thickness of materials is used for leakage current calculations. A temperature of 300 K is assumed.

## 3. Results and discussion

**Fig. 1** shows the side and top views of the atomic structure of the TMNHs under study. More details of the structures such as space group and crystal structure type are listed in **Table S1** in the supplementary information. The layered crystal structures of TMNHs are found in two different types, the *α*-form (TiNBr and TiNCl) with A-A stacking configuration and the *β*-form (HfNBr, HfNCl, ZrNBr, and ZrNCl) with A-B-C stacking configuration [43, 44]. The *α*-form of monolayer TiNBr and TiNCl is a FeOCl-like orthorhombic layered network which is composed of an outer halogen layer (Cl and Br) added to the Ti atoms. The *β*-form, however, is made up of honeycomb-like sheets of M–N layers crammed with halogen atoms. The *β*-form has SmSI (samarium sulfide iodide) layer structure and is more stable in a hostile environment than the *α*-form [44]. The *β*-form structures can stand hot acid solutions whereas the *α*-form structures are hydrolyzed in air.

The monolayer thickness of the TMNHs under study ranges from 7.88 Å (TiNCl) to 9.82 Å (HfNBr and ZrNBr). Structural parameters such as monolayer thickness (t), bulk interlayer distance (d), the percentage difference between d and t (Δ), and lattice constant (a) of all TMNHs under study are reported in **Table S2** in the supplementary information. The optimized monolayer lattice constants of the *β*-form monolayers range from 3.59 Å (HfNCl) to 3.65 Å (ZrNBr), whereas the *α*-form monolayer of TiNCl and TiNBr have lattice constants of 3.26 Å and 3.36 Å respectively. The calculated lattice constant values are in good agreement with experimental and theoretical values reported in other works [16, 18].

First, we calculate the exfoliation energies, $E_{\text{ex}}$, to certify whether materials in this study are layered and can be produced from their bulk crystal counterparts. The exfoliation energy ranges from 15.55 meV/Å$^2$ (TiNCl) to 18.95 meV/Å$^2$ (HfNBr). All $E_{\text{ex}}$ values for TMNH materials are listed in

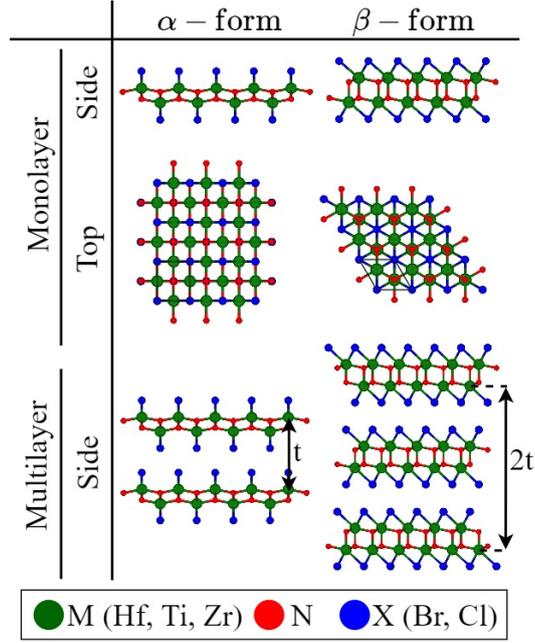

Fig. 1. Structures of the TMNH materials, with chemical formula MNX, under study. The side view and top view of the monolayer in addition to the side view of the multilayer are illustrated. The α-form (A-A stacking configuration) represents TiNBr and TiNCl structures whereas the β-form (A-B-C stacking configuration) represents the structures of HfNBr, HfNCl, ZrNBr, and ZrNCl. The monolayer thickness ($t$) is indicated on the multilayer structures.

**Table S2**. As reported in [45], materials with $E_{ex} < 100$ meV/Å$^2$ are deemed easily exfoliable compounds. According to **Table S2**, all six TMNHs materials in this study have weak vdW interlayer interactions and are therefore exfoliable and layered. Moreover, to examine the stability of monolayers we calculate the phonon energies of monolayer and bulk for all TMNH materials. In the supplementary information, we tabulate the monolayer and bulk form phonon energy values in **Tables S3** and **S4**, respectively. No significant imaginary phonon energies are observed indicating that all monolayer materials are stable.

**Fig. 2** shows the calculated macroscopic in-plane (∥) and out-of-plane (⊥) dielectric constants for the materials under study. The values of the dielectric constants for all materials are listed in **Table 1**. While the optical dielectric constant ($\varepsilon_\infty$) only contains the electronic response, the static dielectric constant ($\varepsilon_0$), on the other hand, includes both the ionic and the electronic contributions to the dielectric response. For the calculation of the dielectric constant of monolayer slabs, we first introduce an appropriate amount of vacuum size (at least 15 Å) to the computational supercell. Applying a sufficient vacuum size is critical to effectively isolate monolayers from other layers and counteract unwanted interlayer interactions. As recently presented in [2, 31], we remove the vacuum effect by rescaling the supercell dielectric constants to those of the monolayer. For the in-plane dielectric calculation, we calculate the geometric mean over the $x$ and $y$ components of dielectric values. In **Table S5** in the supplementary information, the $x$ and $y$ components are separately reported for the monolayer and bulk TMNH materials in this study showing the in-plane anisotropy of TiNBr and TiNCl.

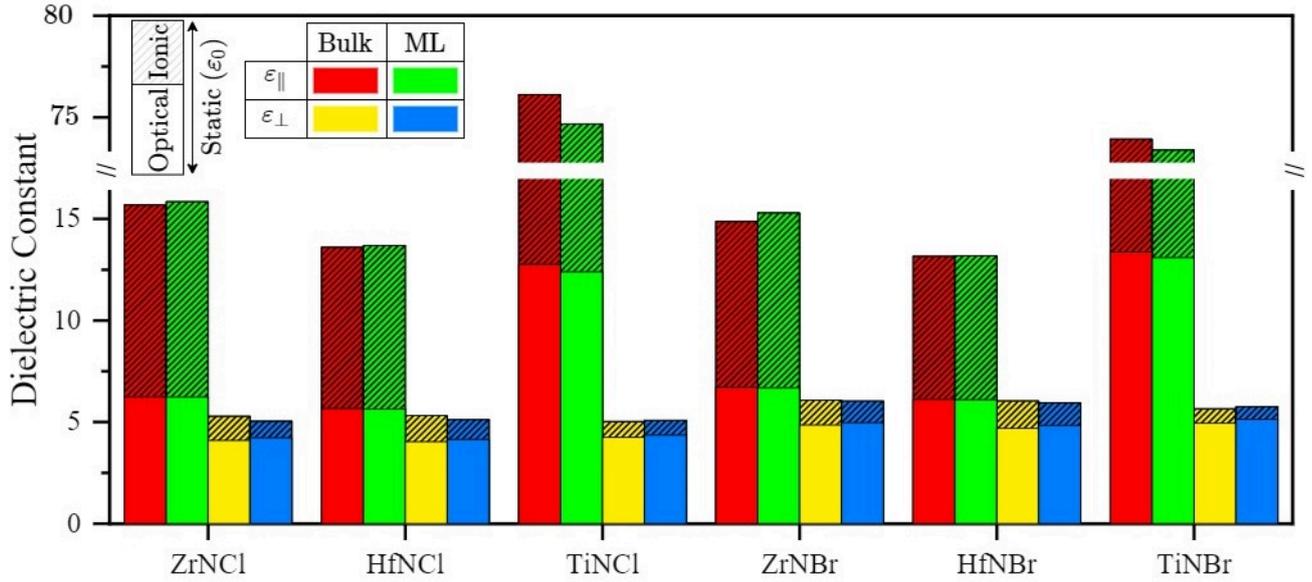

Fig. 2. Dielectric constants for monolayer and bulk of the TMNHs. Both optical ($\varepsilon_\infty$) and static ($\varepsilon_0$) dielectric constants in the in-plane (∥) and out-of-plane (⊥) directions are shown. Larger shaded area represents higher ionic contribution in dielectric constants. TiNBr and TiNCl have relatively high ionic response in their dielectric constants. All dielectric values are reported in Table 1. Only TiNBr and TiNCl are anisotropic materials, $\varepsilon_x \neq \varepsilon_y$, (see Table S5).

Table 1. Dielectric constant values for monolayer and bulk TMNH materials under study. Both optical ($\varepsilon_\infty$) and static ($\varepsilon_0$) dielectric constants in the in-plane (∥) and out-of-plane (⊥) directions are reported.

| Material | Monolayer ($\varepsilon_\infty$) | | Monolayer ($\varepsilon_0$) | | Bulk ($\varepsilon_\infty$) | | Bulk ($\varepsilon_0$) | |
|---|---|---|---|---|---|---|---|---|
| | ∥ | ⊥ | ∥ | ⊥ | ∥ | ⊥ | ∥ | ⊥ |
| ZrNCl | 6.23 | 4.23 | 15.85 | 5.04 | 6.24 | 4.10 | 15.68 | 5.28 |
| HfNCl | 5.65 | 4.15 | 13.69 | 5.13 | 5.67 | 4.04 | 13.61 | 5.32 |
| TiNCl | 12.40 | 4.36 | 74.52 | 5.08 | 12.76 | 4.25 | 75.74 | 5.02 |
| ZrNBr | 6.69 | 4.97 | 15.30 | 6.03 | 6.72 | 4.86 | 14.87 | 6.08 |
| HfNBr | 6.10 | 4.84 | 13.18 | 5.94 | 6.12 | 4.70 | 13.16 | 6.04 |
| TiNBr | 13.10 | 5.14 | 67.01 | 5.76 | 13.37 | 4.95 | 68.22 | 5.65 |

For the bulk samples, **Fig. 1** shows that TiNCl (75.74) and HfNBr (13.16) exhibit the highest and the lowest in-plane static dielectric constants, respectively, whereas ZrNBr (6.08) and TiNCl (5.02) display the highest and the lowest out-of-plane static dielectric constants, respectively. The in-plane static dielectric constants of TiNBr and TiNCl are approximately six times larger than the corresponding optical dielectric constants indicating that there is a significant ionic contribution to the dielectric response for TiNBr and TiNCl. The optical dielectric constants range from 5.67 (HfNCl) to 13.37 (TiNBr) in the in-plane direction and locate between 4.04 (HfNCl) and 4.95 (TiNBr) in the out-of-plane direction, respectively.

In monolayers, **Fig. 1** reveals that the highest and the lowest in-plane static dielectric constants are attributed to TiNCl (74.52) and HfNBr (13.18), respectively. In addition, the out-of-plane static dielectric constants lie between 5.04 for ZrNCl and 6.03 for ZrNBr. Similar to the bulk structures, the ionic contribution of TiNBr and TiNCl to the in-plane dielectric responses are noticeably high. For the optical dielectric constants, HfNCl has the lowest in-plane (5.65) and out-of-plane (4.15) dielectric constants whereas, TiNBr has the highest in-plane (13.10) and out-of-plane (5.14) dielectric constants.

There is a small difference between the dielectric constant values of monolayer and bulk in both in-plane and out-of-plane directions. Going from bulk to monolayer, we observe that the in-plane optical dielectric constants decrease less than 3%, while the out-of-plane optical response increases by up to 4%. Similar to the in-plane optical dielectric constants, the in-plane static dielectric constants show less than 3% difference between bulk and monolayer dielectric constants. However, by moving from bulk to monolayer we detect an increase in the in-plane dielectric constant for ZrNCl, HfNCl, ZrNBr, and HfNBr whereas, the trend is the opposite for TiNBr and TiNCl, such that their dielectric constants decrease marginally. For the out-of-plane static dielectric constants, the maximum difference is exhibited by ZrNCl with an increase of 4.8% going from monolayer to bulk.

**Fig. 3** shows the energy band alignment of the TMNH materials, hBN, and TMDs. We calculate the electron affinity, bandgap, and conduction-valence band edges of TMNHs and TMDs to analyze the insulating properties of TMNHs for transistors with TMD channel materials. We report the bandgap and the electron affinity (difference between the vacuum level and the conduction band minimum) values of the TMNHs and TMDs on **Fig. 3**. We also add hBN, a very well-known 2D gate dielectric to our figure for the purpose of comparison. The green and yellow bars respectively show the conduction, and the valence band positions relative to the vacuum reference, which is set at zero. The monolayer bandgap and electron affinity values of all materials in this study are calculated by Heyd–Scuseria–Ernzerhof (HSE06) hybrid functional method to correct for bandgap underestimation. The underestimated Perdew-Burke-Ernzerof (PBE) values of TMNHs are provided in **Table S6** in the supplementary information.

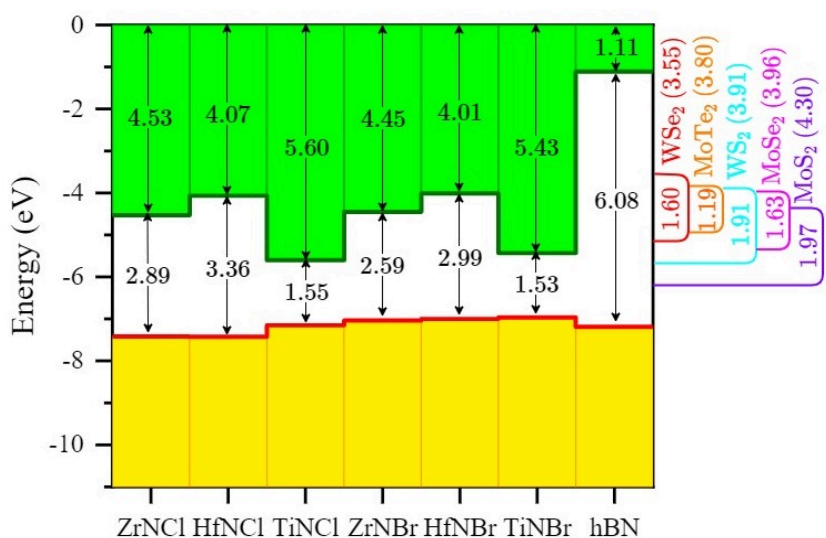

Fig. 3. Band alignment of TMNHs. Top bars (green) and bottom bars (yellow) show conduction band and valence band energies, respectively. The bandgap and electron affinity values of each TMNH are indicated on the figure (Table S6 in the supplementary information). hBN is added for the purpose of comparison (Table S7 in the supplementary information). The band alignment of TMDs, showing the bandgap and electron affinity values (in parenthesis), are placed on the right side of the figure as guides to the eye.

As a general principle, a suitable dielectric candidate should possess a band offset exceeding 1 eV with the channel material to efficiently minimize leakage current[46]. Our calculation shows that none of TMNHs in our list are good candidate dielectrics for *n*-MOS TMD applications because of the small TMNH-TMD offset which would result in an unacceptably high dielectric leakage current. However, there are promising TMNH dielectric materials for *p*-MOS transistors owing to a sufficiently large bandgap and favorable band offset with respect to the TMD channel materials. Furthermore, **Fig. 3** shows that all TMNH materials and hBN have a similar valence band edge and promise performance similar to hBN.

We calculate the EOT values for monolayer and bilayer (2L) of all six TMNH dielectric materials and plot the results in **Fig. 4**. We add monolayer and bilayer hBN, reported in [31], for the purpose of comparison. For the monolayer EOT (narrow bars), we consider the monolayer thickness from **Table S2** and the monolayer out-of-plane static dielectric constant from **Table 1**. For the bilayer EOT (wide bars) we present a range between two different assumptions: one where the bilayer has the dielectric constant of the monolayer and one where the bilayer has the dielectric constant of the bulk. We assume the bilayer thickness equals twice the monolayer thickness. The shaded area for each TMNH shows the range the bilayer EOT lies in, the shaded area is small indicating that both assumptions for bilayer dielectric constant yield similar results.

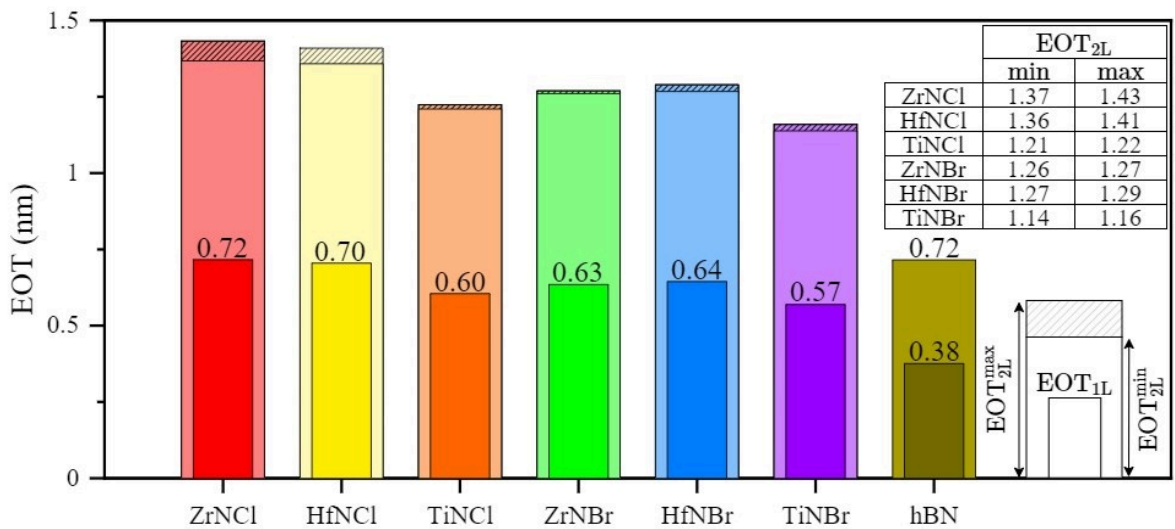

Fig. 4. Monolayer (narrow bars) and bilayer (wide bars) EOT for the TMNHs under study. Monolayer EOT is calculated from monolayer out-of-plane static dielectric constants and monolayer thickness. Bilayer EOT is calculated at two extremes assuming the bilayer dielectric constant either equals the monolayer or the bulk out-of-plane static dielectric constants. In the inset, two extreme values of bilayer EOT for gate dielectrics are reported. hBN values are calculated from the hBN dielectric constant and thickness for monolayer and bilayer, as reported in [31].

Monolayer EOT values of TMNH materials in this study range from 0.57 nm (TiNBr) to 0.72 nm (ZrNCl) as shown in **Fig. 4**. Our results show we can benefit from ultrathin 1-nm EOT in all materials that can subjugate short-channel effects and compete with other 2D layered ultrathin dielectric materials. Taking only EOT into account, monolayer hBN with an EOT of 0.38 nm outperforms the TMNHs investigated here. However, monolayer hBN leakage currents are known to be unacceptable in CMOS technology [47].

We also determine the leakage current density of TMNHs. **Fig. 5** shows the dielectric leakage current of *p*-MOS transistors with the six monolayer TMNHs under study in combination with five well-known TMDs, yielding 30 combinations in total. We add monolayer hBN for the purpose of comparison. The leakage current includes three components: thermionic emission, tunneling current through the valence band, and tunnelling current through the conduction band.

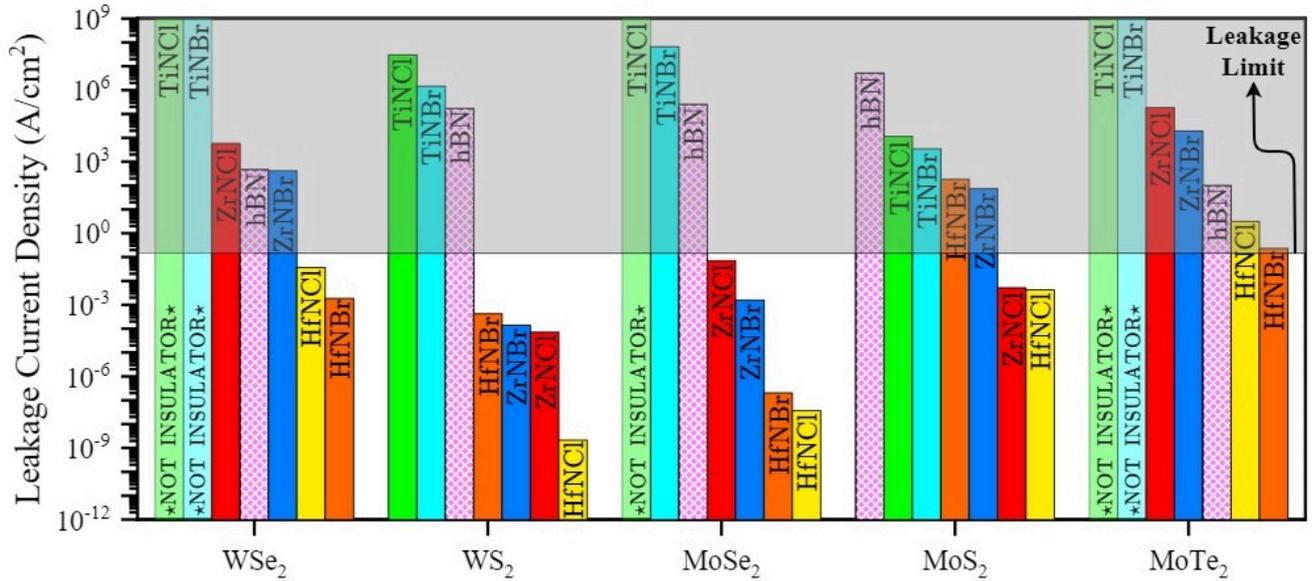

Fig. 5. Leakage current of *p*-MOS transistors with six monolayer TMNHs under study in combination with five well-known TMDs. hBN (pink bar) is added for the purpose of comparison. TiNCl and TiNBr cannot insulate current when they are combined with WSe$_2$ and MoTe$_2$ due to their band alignment (shown in Fig. 3). Moreover, TiNCl cannot insulate the current when it is used with MoSe$_2$. The shaded area indicates unaccepted leakage currents, *i.e.* exceed the leakage current limit according to IRDS [39].

We report the leakage current values for each component as well as the total leakage current values in **Table S8** in the supplementary information. The thermionic current is relatively negligible for all materials whereas the tunneling currents are dominant. More details about the leakage current calculation are provided in the method section. For each TMD channel material, TMNH materials with bars on the right side of the hBN bar outperform the hBN. 19 combinations have a smaller leakage current than hBN. The smallest leakage current of $2.14\times10^{-9}$ A/cm$^2$ belongs to a WS$_2$ *p*-MOS transistor with 2D HfNCl as a dielectric gate. Our calculation shows that TiNCl and TiNBr cannot insulate the current when they are combined with WSe$_2$ and MoTe$_2$ due to their band alignment (shown in **Fig. 3**). In addition, TiNCl is not able to insulate the current when it is coupled with MoSe$_2$.

According to the 2020 IRDS [41], a maximum leakage current of 100 pA/μm needs to be considered for any viable gate dielectric. Assuming a 28 nm pitch, an 18 nm long gate, and an effective gate with of 107 nm, the acceptable leakage current density becomes 0.145 A/cm$^2$. The shaded area on **Fig. 5** shows the unaccepted leakage current density based on IRDS criteria. Among the 30 combinations of TMNHs and TMDs proposed in this study, 12 combinations result in a leakage current smaller than the limit, 0.145 A/cm$^2$. Among the five TMDs, MoSe$_2$ and WS$_2$ are the most promising channel materials in terms of leakage current since their combination with four TMNHs (HfNBr, HfNCl, ZrNBr, and ZrNCl), make *p*-MOS transistors with acceptable leakage currents. On the other hand, the combination of MoTe$_2$ with all six TMNHs appear to result in unacceptable leakage currents. Moreover, the leakage current of TiNBr, TiNCl, and hBN with all five TMDs exceed the leakage current limit.

The excessive leakage current of hBN with all five TMDs, showing in **Fig. 5**, is in good agreement with a recent finding by Knobloch et al. in [47]. Knobloch et al. performed theoretical work, using DFT, and experimental work, using Chemical Vapor Deposition (CVD) and exfoliation of hBN, and reported that hBN is unlikely to be suitable for ultrascaled CMOS transistors due to excessive leakage current. Knobloch et al. used trilayer hBN as a gate insulator on *p*-type silicon and calculated leakage current revealing that the leakage current of devices with hBN exceeds the low-power limit, however, when operating when the voltage drop over the oxide is lower than 0.5 V, hBN trilayers could be used. In another recent study [2], it was shown that hBN leakage current is larger than the leakage current limit determined by IRDS. Finally, we note that recently six monolayer 2D halide dielectrics (HoOI, LaOBr, LaOCl, LaOI, SrI2, and YOBr), suitable for *n*-MOS applications, have been proposed in Ref. [2] to outperform hBN and $HfO_2$.

## 4. Conclusions

We investigated TMNH layered dielectric materials which are classified into two forms (*α*-form and *β*-form) based on their lattice structures and space groups. We carried out first-principles calculations to accurately compute in-plane and out-of-plane optical and static dielectric constants for the bulk and monolayers of six TMNHs. The monolayer in-plane static dielectric constant of monolayer TiNCl and TiNBr were shown to be much greater than the other materials, whereas the out-of-plane static dielectric constant for all materials is in the same range, where ZrNBr has the largest dielectric constant. We calculated phonon and exfoliation energies proving that all TMNH bulk materials are thermodynamically stable and potentially exfoliable down to monolayers. We also estimated monolayer HSE bandgap values, electron affinities, and band edges for TMNH materials and TMD channels.

To investigate the potential of materials under study for *p*-MOS applications, we calculated EOT and leakage current of transistors with TMNHs as gate dielectrics and TMDs as channels materials. All six TMNHs have an EOT < 1 nm suitable for ultrascaled CMOS devices. We found that hBN has excessive leakage current with all five TMDs. Among 30 combinations (six TMNHs and five TMDs), 19 combinations showed a smaller leakage current than hBN while 12 combinations have a leakage current smaller than the low-power limit introduced by IRDS. The combination of monolayer HfNCl with $WS_2$ had the smallest leakage current density ($2.14\times10^{-9}$ A/cm$^2$). $WS_2$ and $MoSe_2$ were found as the most promising TMDs in this study since their combination with four TMNHs (HfNBr, HfNCl, ZrNBr, and ZrNCl) resulted in relatively low leakage currents. TiNBr and TiNCl with all five TMDs result in large leakage currents exceeding the leakage current limit defined by IRDS. Based on our findings, we propose four promising 2D gate dielectrics (HfNBr, HfNCl, ZrNBr, and ZrNCl) suitable for ultrascaled TMD *p*-MOS transistors. Our approach can be used to discover more 2D gate dielectric materials and investigate the performance of novel ultrascaled CMOS devices.

## Acknowledgements

This research was sponsored in part by the Semiconductor Research Corporation (SRC) under the Logic and Memory Devices (LMD) Program of the Global Research Collaboration (GRC).